\documentclass[%
reprint,
superscriptaddress,
groupedaddress,
frontmatterverbose,
showpacs,preprintnumbers,
amsmath,amssymb,
aps,
prl,
]{revtex4-1}

\usepackage{mathtools}
\usepackage{graphicx}
\usepackage{dcolumn}
\usepackage{bm}
\usepackage{verbatim}
\usepackage{hyperref}
\usepackage{amsmath}
\usepackage{braket}

\usepackage[utf8]{inputenc}
\usepackage{graphicx}
\usepackage{dcolumn}
\usepackage{bm}
\usepackage{booktabs}
\usepackage{amsmath}
\usepackage{amssymb}
\usepackage{nicefrac}
\usepackage{mathrsfs}
\usepackage{braket}
\usepackage{subfigure}
\usepackage{blindtext}
\usepackage{enumerate}
\usepackage{setspace}
\usepackage{tabularx}

\usepackage{mathtools}
\usepackage{graphicx}
\usepackage{dcolumn}
\usepackage{bm}
\usepackage{verbatim}
\usepackage{hyperref}
\usepackage{amsmath}

\newcommand{\erz}[2]{\hat{\operatorname{#1}}_{#2}^{\dag}}			
\newcommand{\ver}[2]{\hat{\operatorname{#1}}_{#2}^{\phantom{\dag}}}		
\newcommand{\kk}{\mathbf{k}}							
\newcommand{\HH}[1]{\hat{H}_{\mathrm{#1}}}		  	
\newcommand{\eps}[1]{\epsilon_{#1}}						
	
\newcommand{\ee}[1]{\mathrm{e}^{#1}}						
\newcommand{\mc}[1]{\multicolumn{1}{c}{#1}}					
\renewcommand{\em}{\it}

\AtBeginDocument{
\heavyrulewidth=.08em
\lightrulewidth=.05em
\cmidrulewidth=.03em
\belowrulesep=.65ex
\belowbottomsep=0pt
\aboverulesep=.4ex
\abovetopsep=0pt
\cmidrulesep=\doublerulesep
\cmidrulekern=.5em
\defaultaddspace=.5em
}

\begin{document}

\title{Apparent Reversal of Molecular Orbitals Reveals Entanglement}

\author{Ping Yu}
\affiliation{Institute of Experimental and Applied Physics, University of Regensburg, 93053 Regensburg, Germany}
\affiliation{School of Physical Science and Technology, ShanghaiTech University}
\author{Nemanja Koci\'{c}}
\affiliation{Institute of Experimental and Applied Physics, University of Regensburg, 93053 Regensburg, Germany}
\author{Benjamin Siegert}
\affiliation{Institute of Theoretical Physics, University of Regensburg, 93053 Regensburg, Germany}
\author{Andrea Donarini}
\email{andrea.donarini@ur.de}
\affiliation {Institute of Theoretical Physics, University of Regensburg, 93053 Regensburg, Germany}
\author{Jascha Repp}
\email{jascha.repp@ur.de}
\affiliation{Institute of Experimental and Applied Physics, University of Regensburg, 93053 Regensburg, Germany}

\renewcommand{\em}{\it}

\date{\today}

\begin{abstract}
The frontier orbital sequence of individual dicyanovinyl-substituted oligothiophene molecules is studied by means of scanning tunneling microscopy.
On NaCl/\- Cu(111)  the molecules are neutral and the two lowest unoccupied molecular states
are observed in the expected order of increasing energy.
On
NaCl/Cu(311), where
the molecules are negatively charged,
the sequence of two observed molecular orbitals is reversed, such that the one with one more 
nodal plane appears lower in energy.
These experimental results, in open contradiction with a single-particle interpretation, are explained by a many-body theory predicting a strongly entangled doubly charged ground state.
\end{abstract}

\pacs{68.37.Ef, 68.43.-h, 73.23.Hk}        


\maketitle

For the use of single molecules as devices, engineering and control of their intrinsic electronic properties is all-important.
In this context, quantum effects such as electronic interference have recently shifted into 
the focus~\cite{Cardamone2006,Donarini2009,Donarini2010,Guedon2012,Vazquez2012,Ballmann2012,Xia2014}.
Most intriguing in this respect are electron correlation effects~\cite{Zhao2005,Maruccio2007,Fernandez2008,Franke2011,Chiesa2013,Grothe2013,Ervasti2016},
which are intrinsically strong in molecules due to their small size~\cite{Begemann2008,Toroz2011,Toroz2013,Schulz2015,Siegert2016}.

In general, Coulomb charging energies strongly depend on the localization of electrons and hence on the spatial extent of the orbitals they occupy. Therefore 
the orbital sequence of a given molecule can reverse upon electron attachment or removal, if some of the frontier orbitals are strongly localized while others are not, 
like in e.\,g. phthalocyanines~\cite{Liao2001,Nguyen2003,Wu2006,wu2008,Uhlmann2013}.
Coulomb interaction may also lead to much more complex manifestations such as 
quantum entanglement of delocalized molecular orbitals.

Here we show, that the energy spacing of the frontier orbitals in a single molecular wire of individual dicyanovinyl-substituted quinquethiophene (DCV5T) can be engineered to achieve
near-degeneracy of the two lowest lying unoccupied molecular orbitals, leading to a strongly-entangled ground state of DCV5T$^{2-}$. 
These orbitals are the lowest two of a set of  particle-in-a-box-like states and differ only by one additional nodal 
plane across the center of the wire. Hence, according to the fundamental oscillation theorem of the Sturm-Liouville theory their sequence 
has to be set with increasing number of nodal planes,
which is one of the basic principles of quantum mechanics~\cite{Hueckel1931,Barford2005}.
This is evidenced and visualized from scanning tunneling microscopy
(STM) and spectroscopy (STS) of DCV5T on ultrathin insulating films. Upon lowering the substrate's work function, the molecule becomes charged, leading to a reversal of 
the sequence of the two orbitals. The fundamental oscillation theorem seems strikingly violated since the state with one {\em more} nodal plane appears {\em lower} in energy.
This contradiction can be solved, though, by considering intramolecular correlation leading to a strong entanglement in the ground state of DCV5T$^{2-}$.

The experiments were carried out with a home-built combined STM/atomic force microscopy (AFM) using a qPlus sensor \cite{giessibl2000} operated in ultra-high vacuum
at a temperature of $6$~K. 
Bias voltages are applied to the sample. All AFM data, d$I$/d$V$ spectra and maps were acquired in constant-height mode. 
Calculations of the orbitals and effective single particle electronic structure were performed within the density functional theory (DFT) as implemented in the SIESTA code~\cite{soler2002} and are based on the generalized gradient approximation (GGA-PBE). 
The many-body eigenstates are determined from a diagonalization of the many-body model Hamiltonian $H_{\rm mol}$, which is defined further below in the main text. Based on these, STM-image and spectra simulations were performed within a Liouville approach for the density matrix $\rho$. 
See Supplemental Material~\cite{Suppl} for more details.

\begin{figure}
  \centering
  \includegraphics[width=\columnwidth]{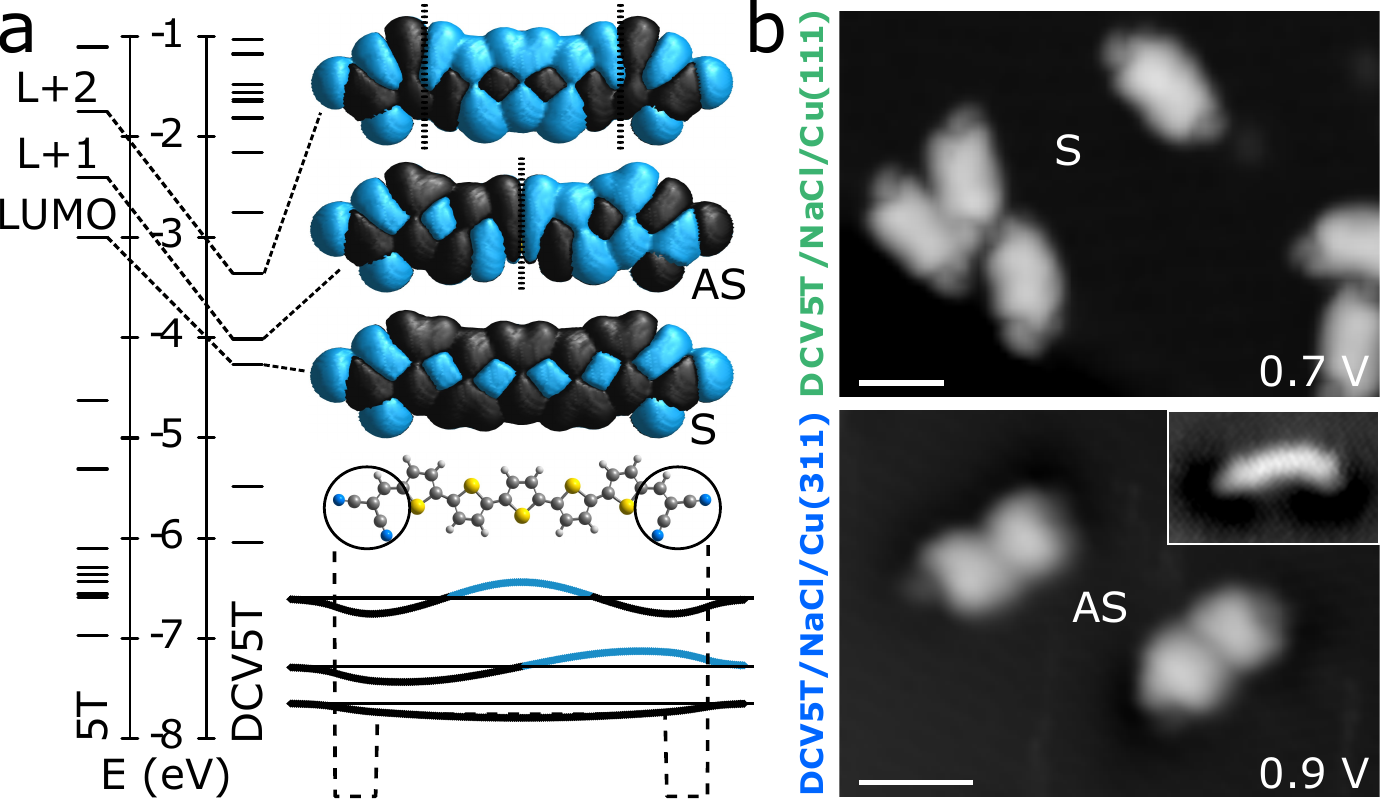}
  \caption{(a) Molecular structure and density-functional theory based calculations of the electronic structure of 5T and DCV5T. 
The panel depicts the molecular structure, the calculated orbitals and energies for the LUMO, LUMO+1 and LUMO+2 as indicated. The orbitals are depicted as contours of constant probability density. 
The LUMO and LUMO+1 orbitals 
derive from the thiophene subunit's LUMO. They are the lowest two of a set of particle-in-a-box-like states and differ only by one additional nodal plane. Whereas the LUMO to LUMO+1 energy difference is approx. 0.7\,eV for 5T, this difference is drastically reduced in the case of DCV5T. The basic principle of level engineering is illustrated for a one dimensional quantum box. (b) STM images of the first DCV5T electronic resonance for NaCl/Cu(111) (top) and NaCl/Cu(311) (bottom) as substrates. Insets show corresponding STM images at voltages below the first molecular resonance.
}
    \label{Fig:DFT}
\end{figure}

The molecular structure of DCV5T, shown in Fig.~\ref{Fig:DFT}a, consists of a quinquethiophene (5T) backbone and a dicyanovinyl (DCV) moiety at each end. 
The delocalized electronic system of polythiophene and oligo-thiophene enables conductance of this material~\cite{Feast1996,Yamada2008,Fitzner2011}.
The lowest unoccupied orbital of each of the thiophene rings couples electronically to its neighbors and forms a set of particle-in-a-box-like states~\cite{Repp2010,Kislitsyn2016}.
The LUMO to LUMO+1 level spacing of the quinquethiophene (5T) backbone is approx. 0.7~eV~\cite{Repp2010}, which is in 
good agreement with the energy difference calculated for free 5T based on DFT, as shown in Fig.~\ref{Fig:DFT}a, left. This 
DFT-based calculation also confirms the nature of the LUMO and LUMO+1 orbitals, both deriving from the single thiophene's LUMOs and essentially differing only by one 
additional nodal plane across the center of the molecule. To enable the emergence of correlation and thus level reordering, we have to bring these two states closer to each other. 
This is achieved by substituting dicyanovinyl moieties with larger electron affinity at each end of the molecular wire. 
As the orbital density of the higher lying particle-in-a-box-like state, namely LUMO+1,  has more weight at the ends of the molecule, 
it is more affected by this substitution than the lowest state, the LUMO. This is evidenced by corresponding calculations of DCV5T, for which the LUMO to LUMO+1 energy difference is reduced by more than a factor of two, see Fig.~\ref{Fig:DFT}a, left. The increased size of DCV5T may also contribute to the reduced level spacing. 
For the rest of this work, we concentrate on the LUMO and LUMO+1 orbitals only. To avoid confusion, we refrain from labeling the orbitals according to their sequence but instead 
according to their symmetry with respect to the mirror plane perpendicular to the molecular axis, as symmetric (S) and antisymmetric (AS). Hence, the former LUMO and the LUMO+1 are the S and AS states, respectively.

To study the energetic alignment of the orbitals as well as their distribution in real space, we
employ ultrathin NaCl insulating films to electronically decouple the molecules from the conductive substrate~\cite{Repp2005}. It has been previously shown that in these systems the work function can be changed by using different surface orientations of the underlying metal support~\cite{Repp2005,Olsson2007,Swart2011}. Importantly, this does not affect the 
(100)-terminated surface orientation of the NaCl film, such that the local chemical environment of the molecule remains the same, except for the change of the work function.

%
%

However, in the present case, this alone has a dramatic effect on the electronic structure of the molecular wires as is evidenced in Fig.~\ref{Fig:DFT}b. There, the STM images are shown for voltages corresponding to the respective lowest lying molecular resonances at positive sample voltage for DCV5T adsorbed on NaCl/Cu(111) (top panel) and NaCl/Cu(311) (bottom panel). They both show a hot-dog like appearance of the overall orbital density as was observed and discussed previously~\cite{Repp2010,Bogner2015}. Importantly in the current context, however, the orbital density of DCV5T/NaCl/Cu(311) shows a clear depression at the center of the molecule, indicating a nodal plane, whereas DCV5T/NaCl/Cu(111) does not. Apparently, the energetically lowest lying state is not the same for the two cases, but S for DCV5T/NaCl/Cu(111) and AS in the case of DCV5T/NaCl/Cu(311). In contrast, STM images acquired at voltages well below the first resonance reflect the geometry of the molecule in both cases as wire-like protrusion (see insets of Fig.~\ref{Fig:DFT}b).


We hence assume that the molecules are neutral on NaCl/Cu(111) and that the S state corresponds to the LUMO. According to the literature, changing the copper surface orientation from Cu(111) to Cu(311) results in a lowering of the work function by approximately $1$\,eV ~\cite{Gartland1972,Repp2005,Olsson2005}. Hence, one may expect that the former LUMO, initially located $0.7$\,eV above the Fermi level $E_F$ in the case of NaCl/Cu(111) will shift to below the Fermi level~\cite{Swart2011,Uhlmann2013} for NaCl/Cu(311) such that the molecule becomes permanently charged.



\begin{figure}
  \centering
  \includegraphics[width=\columnwidth]{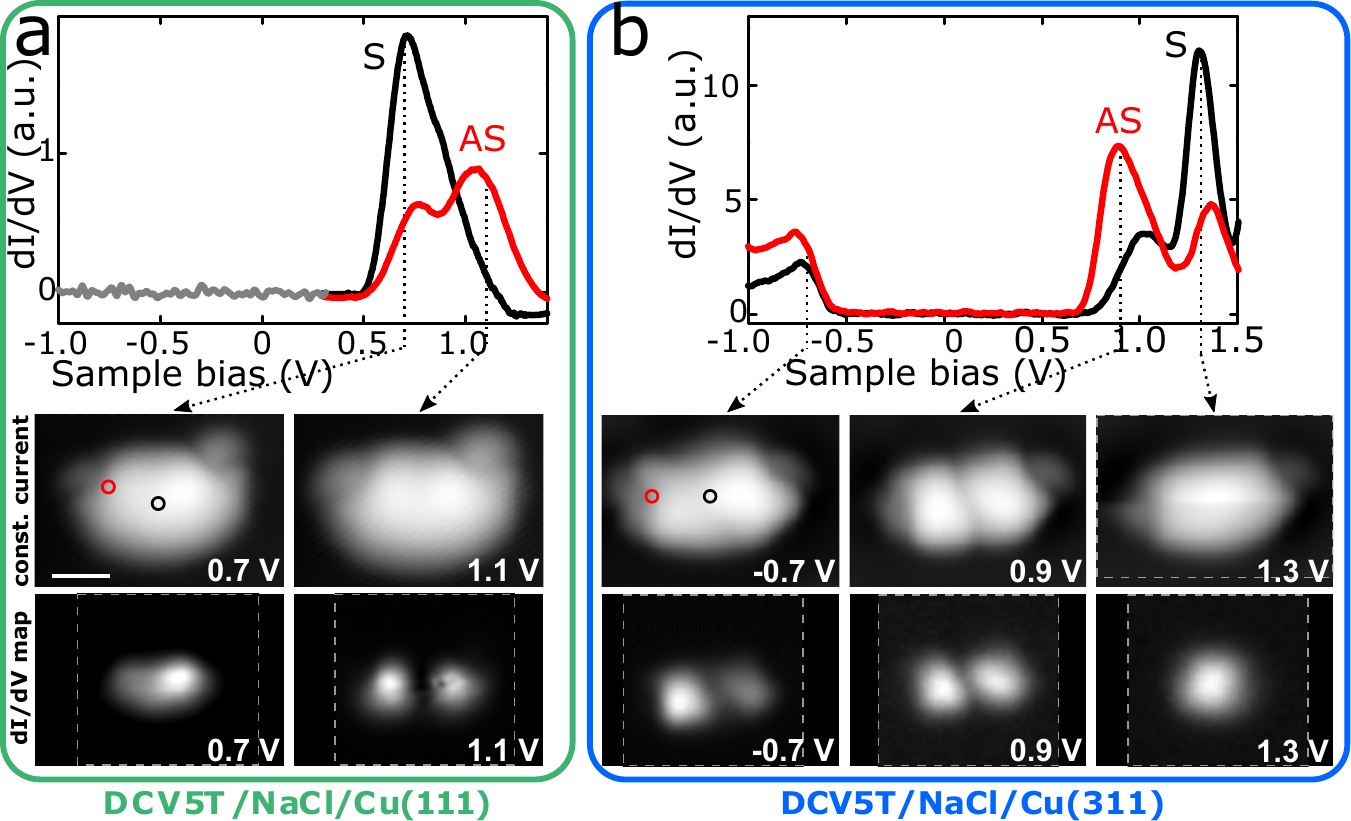}\\
  \caption{d$I$/d$V$ spectra (top panels), constant-current STM images (center panels) and d$I$/d$V$ maps (bottom panels) on the individual molecule DCV5T on NaCl/Cu(111) (a) and NaCl/Cu(311) (b) respectively. The resonances are labelled with S and AS, referring to the symmetic and antisymmetric states, respectively.  d$I$/d$V$ spectra were recorded on (black) and off (red) the center of the molecule as indicated by dots in the constant-current STM images, using lock-in detection. To not miss any small d$I$/d$V$ signals in the low-bias range, a corresponding spectrum (grey) was measured at different set-point values 
such that the tip was $\simeq 2$\,\AA\ closer to the surface compared to the other two spectra (red and black). All spectra were slightly low-pass filtered. 
The images are resized to have the same size and scale, whereby the area of measured data is indicated with white dashed rectangles. Constant current images $I = 2.4$\,pA; bias voltage as indicated. Scale bar 1 nm.}
  \label{Fig:Experiments}
\end{figure}

To obtain a systematic understanding of the level alignment of the S and AS states of the molecule on both substrates, we acquired differential conductance (d$I$/d$V$) spectra and d$I$/d$V$-maps on DCV5T molecules. Typical spectra measured at the center and the side of the molecule are shown in Figs.~\ref{Fig:Experiments}a and b on NaCl/Cu(111) and NaCl/Cu(311), respectively. DCV5T exhibits two d$I$/d$V$ resonances at positive bias but none at negative voltages down to -2.5~V. According to the d$I$/d$V$ maps and consistent with the different intensities in the spectra acquired on and off center of the molecule, the S state at $\simeq0.7$~V is lower in energy than the AS state occurring at $\simeq1.1$~V. The energy difference of $\simeq0.4$~eV is in rough agreement to our calculations (see Fig.~\ref{Fig:DFT}a).
 As discussed above, in the case of NaCl/Cu(311), DCV5T exhibits the AS state as the lowest resonance at positive bias voltages, this time at $\simeq 0.9$~V. This is additionally evidenced by the constant-current STM image and the corresponding d$I$/d$V$ map in Fig.~\ref{Fig:Experiments}b. The S state is now located at higher voltages, namely at $\simeq1.3$~V, as seen in the spectrum and the d$I$/d$V$ map. Obviously, the two states are reversed in their sequence. In this case, at negative bias voltages, a peak in d$I$/d$V$ indicates an occupied state in equilibrium, in stark contrast to DCV5T/NaCl/Cu(111) but in agreement with the assumption of the molecule being negatively charged. The constant-current image acquired at $-0.7$~V, corresponding to the first peak at negative bias, seems to be a superposition of both the S and AS states.

\begin{figure}
  \centering
  \includegraphics[width= 8 cm]{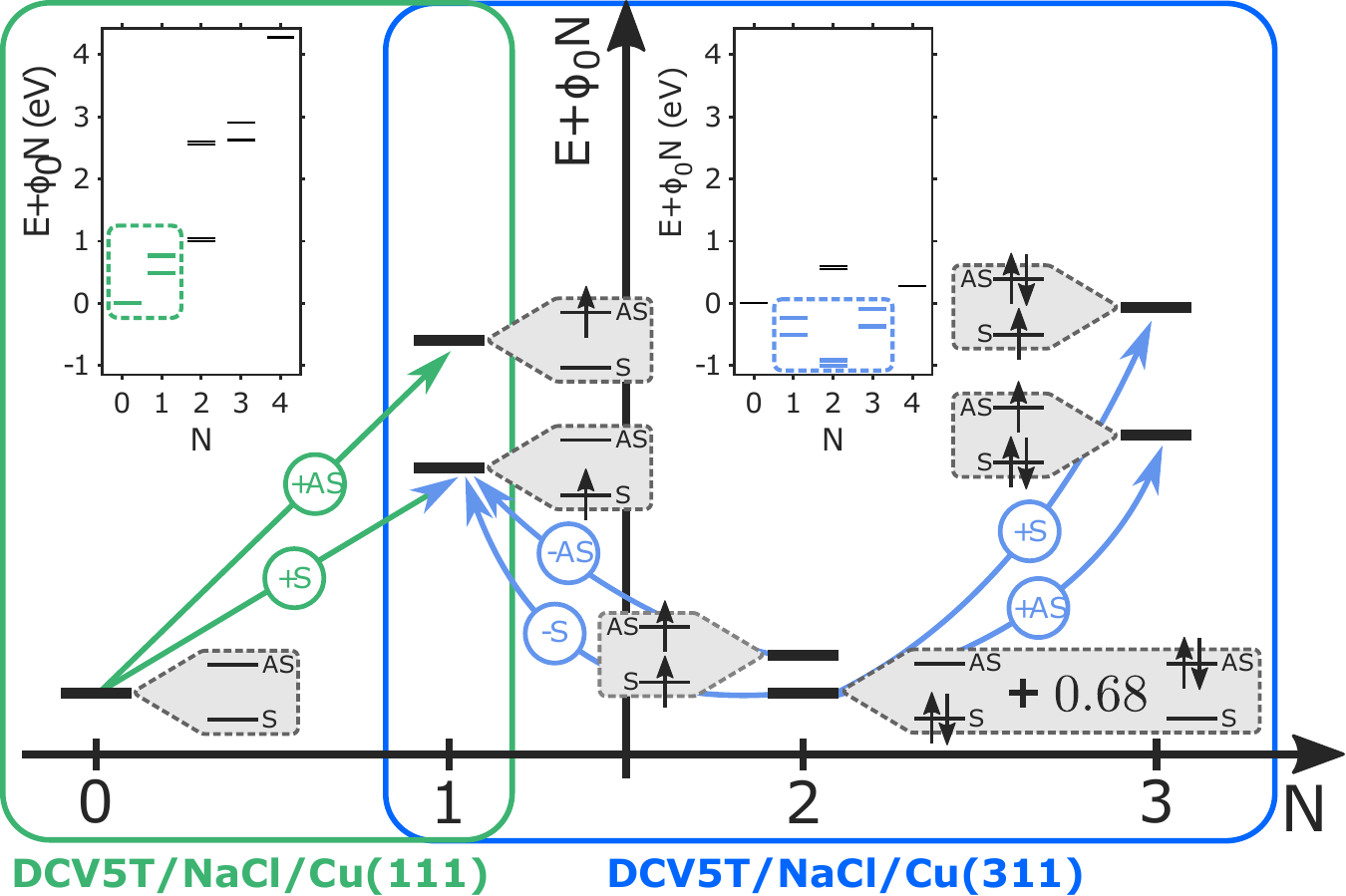}\\
  \caption{Scheme with the many-body transitions associated to the measured resonances.
  In the green framed panel the transition between the neutral and the singly charged DCV5T$^-$ are illustrated (DCV5T/NaCl(Cu(111)). In the blue framed panel the transitions involving DCV5T$^{-}$, DCV5T$^{2-}$ and DCV5T$^{3-}$ are analyzed (DCV5T/NaCl(Cu(311)). The electronic structure associated to the different many-body states is explicitly given in the gray labels. In the insets, the many-body spectra of the molecule on the two corresponding substrates are plotted.}
  \label{Fig:Scheme}
\end{figure}

The experimentally observed reversal of the orbital sequence is in striking disagreement with the fundamental oscillation theorem. To understand this apparent orbital reversal we go beyond the single particle picture and invoke the role of electronic correlations. In the double-barrier tunneling junction geometry employed here, the resonances in d$I$/d$V$ are associated with a temporary change of electron number on the molecule. In this terms the two peaks of DCV5T/NaCl/Cu(111) at positive bias are DCV5T$\leftrightarrow $DCV5T$^-$ \emph{transitions} (See Fig.~\ref{Fig:Scheme}), and, in the same spirit, the ones of DCV5T/NaCl/Cu(311) at positive and at negative bias should be interpreted as 
DCV5T$^{2-}\leftrightarrow $DCV5T$^{3-}$ and DCV5T$^{2-}\leftrightarrow $DCV5T$^{-}$ transitions, respectively.

Both the topographical and the spectroscopic data presented so far suggest that the electronic transport through DCV5T involves, in the present bias and work function ranges, only the symmetric (S) and the antisymmetric (AS) orbitals. We concentrate on
them and freeze the occupation of the other lower (higher) energy orbitals to 2 (0).
In terms of these S and AS \emph{frontier} orbitals we write the minimal interacting Hamiltonian for the isolated molecule,
\begin{align}\label{H_mol}
\begin{split}
H_{\mathrm{mol}} =& \epsilon_{\mathrm{S}}\hat{n}_{\mathrm{S}} + \epsilon_{\mathrm{AS}}\hat{n}_{\mathrm{AS}} + \frac{U}{2}\hat{N}(\hat{N}-1)\\
+& J\sum_{\sigma \sigma'} d^\dagger_{\mathrm{AS} \sigma} d^\dagger_{\mathrm{S} \sigma'} d^{}_{\mathrm{AS} \sigma'} d^{}_{\mathrm{S} \sigma}\\
+& J\left( d^\dagger_{\mathrm{AS} \uparrow} d^\dagger_{\mathrm{AS} \downarrow} d^{}_{\mathrm{S} \downarrow} d^{}_{\mathrm{S} \uparrow}
+ d^\dagger_{\mathrm{S}\uparrow} d^\dagger_{\mathrm{S} \downarrow} d^{}_{\mathrm{AS} \downarrow} d^{}_{\mathrm{AS} \uparrow}\right),	
\end{split}
\end{align}
where $d^\dagger_{\mathrm{S(AS)} \sigma}$ creates an electron with spin $\sigma$ in the symmetric (antisymmetric)
orbital, $\hat{n}_{\mathrm{i}}$ counts the number of electron in the orbital with ${\mathrm{i}} = {\mathrm{S,AS}}$ and
$\hat{N}$ represents the total number of electrons occupying the two frontier orbitals. The interaction parameters $U = 1.4$~eV and $J = 0.75$~eV are obtained from the DFT orbitals by direct calculation of the associated Coulomb integrals and assuming a dielectric constant $\epsilon_r = 2$ which accounts for the screening introduced by the underlying frozen orbitals \cite{Ryndyk2013,Siegert2016}. As expected from their similar (de-)localization, the Coulomb integrals of the S and AS states are almost identical~\footnote{For the Coulomb integrals we obtain $U_{\mathrm{S-S}}  = 1.37\,$eV, $U_{\mathrm{AS-AS}} = 1.43\,$eV, $U_{\mathrm{S-AS}}= 1.37\,$eV}. Besides a constant interaction charging energy $U$, the
model defined in Eq.~(\ref{H_mol}) contains exchange interaction and pair-hopping terms, both proportional to $J$, which are responsible for the electronic correlation. The electrostatic interaction with the substrate is known to stabilize charges on atoms and molecules \cite{Kaasbjerg2008,Kaasbjerg2011,Olsson2007} due to image charge and polaron formation.
We account for this stabilization with the additional Hamiltonian $H_{\mathrm{mol-env}} = -\delta\hat{N}^2$. The orbital energies
$\epsilon_{\mathrm{S}} = -3.1\,$eV and $\epsilon_{\mathrm{AS}} = -2.8\,$eV as well as the image-charge renormalization $\delta = 0.43\,$eV are obtained from the experimental resonances of the neutral molecule and previous experimental results on other molecules~\cite{Suppl}


Many-body interaction manifests itself most strikingly for the ground state DCV5T$^{2-}$,
which will therefore be discussed at first.
Consider the two many-body states, in which the two extra electrons both occupy
either the S or the AS state:
They differ in energy by the energy $2\Delta$, where
$\Delta = \epsilon_{\mathrm{AS}} - \epsilon_{\mathrm{S}}$ is the single-particle level spacing between the S and the AS state.
These two many-body states interact via pair-hopping of strength $J$,
leading to a level repulsion.
As long as $\Delta \gg J$, this effect is negligible.
In DCV5T, though, the single-particle level spacing $\Delta$ is small compared to the
pair-hopping $J$, leading to an entangled ground state of DCV5T$^{2-}$ as
\begin{align}
\label{Charging_E}
\begin{split}
|2,0 \rangle &= \cos\theta\,d_{\mathrm{S}\uparrow}^\dagger d_{\mathrm{S}\downarrow}^\dagger |\Omega\rangle
+ \sin\theta\,d_{\mathrm{AS}\uparrow}^\dagger d_{\mathrm{AS}\downarrow}^\dagger|\Omega\rangle, 
\end{split}
\end{align}
with $\quad \theta = \frac{\arctan \left(J/\Delta\right)}{2}$ and where $|\Omega\rangle$ is the ground state of neutral DCV5T.
Note that here, as $J/\Delta \approx 2.6$, this state
shows more than 30\% contribution from both constituent states, is strongly entangled,
and therefore it can not be approximated by a single Slater determinant.
The first excited state of DCV5T$^{2-}$ is a triplet with one electron in the
S and one in the AS orbital at about $54\,$meV above the ground state, as shown in Fig.~\ref{Fig:Scheme}.

The level repulsion in DCV5T$^{2-}$ mentioned above leads to a significant reduction of the ground state energy by roughly 0.5\,eV. 
This effect enhances the stability of the doubly charged molecule  to the disadvantage of DCV5T$^{-}$, which has just a single extra electron and therefore does not feature many-body effects.
%
%


Within the framework of the many-body theory, as sketched in Fig.~\ref{Fig:Scheme}, the apparent orbital reversal between Fig.~\ref{Fig:Experiments}a and Fig.~\ref{Fig:Experiments}b is naturally explained. To this end, as mentioned above, tunneling events in the STM experiments have to be considered as {\em transitions} between the many-body states of different
charges $N$ (see arrows in Fig.~\ref{Fig:Scheme}). The spatial fingerprints of the transitions and hence their appearance in STM images is given by the orbital occupation {\em difference} between the two many-body states and is indicated by the labels S and AS in Fig.~\ref{Fig:Scheme}.
When on NaCl(2ML)/Cu(111), the DCV5T molecule
is in its neutral ground state, see green panel in Fig.~\ref{Fig:Scheme}.
A sufficiently large positive sample bias triggers transitions to the singly charged DCV5T$^-$: The S and AS transitions subsequently
become energetically available in the expected order of the corresponding single-particle states.
A fast tunnelling of the extra electron to the substrate restores the initial condition enabling a steady-state current.

When on NaCl(2ML)/Cu(311) the molecule is doubly charged and in the entangled ground state described by Eq.~\eqref{Charging_E}, see Fig.~\ref{Fig:Scheme}. At sufficiently high positive sample bias the transitions to DCV5T$^{3-}$ are opening, enabling electron tunnelling from the tip to the molecule. The topography of these transitions is again obtained by comparing the 2 and the 3 (excess) electron states of DCV5T (cf. Fig.~\ref{Fig:Scheme}). The transition to the 3 particle {\em ground} state occurs by the population of the AS state and it involves the \emph{first} component of the entangled 2 electron ground state only. The second component cannot contribute to this transition, which is bound to involve only a {\em single} electron tunneling event. Correspondingly, at a larger bias the \emph{first excited} 3 particle state becomes accessible, via a transition involving the \emph{second} component of the 2 particle ground state only. This transition has a characteristic S state topography. Hence, although the electronic structure of the 3 electron states does follow the Aufbau principle, the entanglement of the 2 particle ground state leads to the apparent reversal of the orbital sequence.

\begin{figure}
  \centering
  \includegraphics[width=\columnwidth]{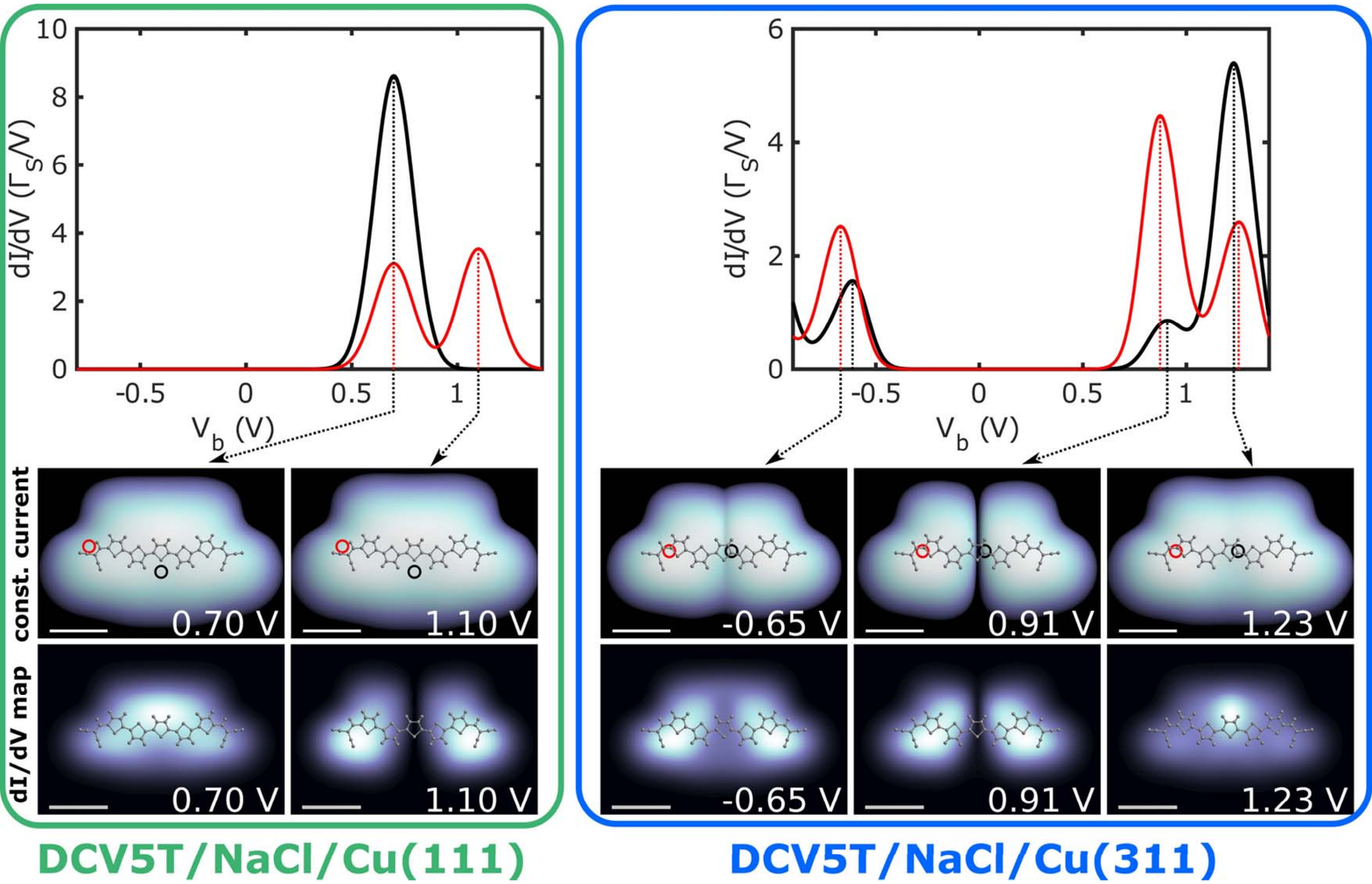}\\
  \caption{Theoretical simulations of d$I$/d$V$ spectra (top panels), constant-current STM images (center panels) and d$I$/d$V$ maps
  (bottom panels) on the individual molecule DCV5T on NaCl/Cu(111) (a) and NaCl/Cu(311) (b) respectively. d$I$/d$V$ spectra were recorded on (black) and off (red) center of the molecule as indicated by dots in the constant-current STM images.
}
  \label{Fig:Simulations}
\end{figure}

As described in the SI, in addition to the many-body spectrum we calculated the full dynamics of subsequent tunneling processes for all relevant situations, resulting in the calculated d$I$/d$V$ characteristics, constant current maps and constant height d$I$/d$V$ maps for a DCV5T single molecule junction presented in Fig.~\ref{Fig:Simulations}. A qualitative agreement with the experimental results of Fig.~\ref{Fig:Experiments} can be observed both for the relative strength of the spectral peaks and the dI/dV maps. The above discussed apparent orbital reversal is fully consistent with the calculations.

The experimental data of DCV5T on the Cu(311) substrate at negative bias also show a non-standard feature. The d$I$/d$V$ map at resonance resembles a superposition of the S and AS orbital, see Fig.~\ref{Fig:Experiments}b. The effect is also reproduced in the theoretical simulations presented in Fig.~\ref{Fig:Simulations}. This can be rationalized in terms of a non-equilibrium dynamics associated to a population inversion recently predicted by some of the authors~\cite{Siegert2016}. 

In conclusion, we showed that a reduction of the single-particle level spacing of two frontier orbitals enables the manifestation of strong electron-correlation effects in single molecules.
Here, the single-particle level spacing engineered by dicyanovinyl-substitution is leading to an apparent reversal of orbital sequence and a strongly-entangled ground state of DCV5T$^{2-}$. The many body description of the electronic transport is capable to reconcile the experimental observations of the orbital reversal with the fundamental oscillation theorem of quantum mechanics and shows how to achieve quantum entanglement of frontier orbitals in molecules. 

\begin{acknowledgments}

The authors thank Milena Grifoni for valuable comments and discussions and David Kasipovi\'{c} for help. Financial support from the Deutsche Forschungsgemeinschaft via SFB 689 and GRK 1570, and the Volks\-wa\-gen Foundation through its Lichtenberg program, are gratefully acknowledged.

\end{acknowledgments}

 \bibliographystyle{apsrev4-1}
 \bibliography{ref}


\onecolumngrid
\appendix
\section{Parametrization of the many-body Hamiltonian}

\noindent The grandcanonical many-body Hamiltonian used in this work,
\begin{align}\label{Eq:H_grand}
	\HH{G} 	= \HH{mol} -\delta\hat{N}^2 + \phi_0\hat{N}
\end{align}
where
 \begin{align}
 	\HH{mol} =& \epsilon_\mathrm{S}\hat{n}_\mathrm{S} + \epsilon_\mathrm{AS}\hat{n}_\mathrm{AS} + \frac{U}{2}\hat{N}(\hat{N}-1)\nonumber\\
 +& J\sum_{\sigma \sigma'} \hat{d}^\dagger_{\mathrm{AS} \sigma} \hat{d}^\dagger_{\mathrm{S} \sigma'} \hat{d}^{}_{\mathrm{AS} \sigma'} \hat{d}^{}_{\mathrm{S} \sigma}\nonumber\\
 +& J\left( \hat{d}^\dagger_{\mathrm{AS} \uparrow} \hat{d}^\dagger_{\mathrm{AS} \downarrow} \hat{d}^{}_{\mathrm{S} \downarrow} \hat{d}^{}_{\mathrm{S} \uparrow}
 + \hat{d}^\dagger_{\mathrm{S} \uparrow} \hat{d}^\dagger_{\mathrm{S} \downarrow} \hat{d}^{}_{\mathrm{AS} \downarrow} \hat{d}^{}_{\mathrm{AS} \uparrow}\right),
 \end{align}
is characterized by six parameters: {\it i.e.} the single particle energies $\epsilon_\mathrm{S}$ and $\epsilon_\mathrm{AS}$ of the frontier orbitals,
the direct interaction and exchange integrals $U$ and $J$, the image charge and polaron renormalization energy $\delta$, the substrate work function $\phi_0$.
Some of these parameters ($U$ and $J$) are obtained from first principle calculations, others  ($\epsilon_\mathrm{S}$ and $\epsilon_\mathrm{AS}$ and $\delta$) are fitted to the present experimental data or taken from the literature ($\phi_0$) \cite{Olsson2005}.

\noindent The direct interaction parameter results from a simplification of a more general model in which all possible combinations of density-density interaction
including the symmetric and the antisymmetric orbital are taken into account. In the most general case one should consider the three parameters:
\begin{equation}
\label{Eq:dir_Interaction_integrals}
\begin{split}
U_{\rm S} &= \frac{e^2}{4\pi \epsilon_r \epsilon_0}\int {\mathrm d}{\bf r}_1 \!\int{\mathrm d}{\bf r}_2 \,\frac{|\psi_{\rm S}({\bf r}_1)|^2 |\psi_{\rm S}({\bf r}_2)|^2}{|{\bf r}_1 - {\bf r}_2|},\\
U_{\rm SAS} &= \frac{e^2}{4\pi \epsilon_r \epsilon_0}\int {\mathrm d}{\bf r}_1 \!\int{\mathrm d}{\bf r}_2 \,\frac{|\psi_{\rm S}({\bf r}_1)|^2 |\psi_{\rm AS}({\bf r}_2)|^2}{|{\bf r}_1 - {\bf r}_2|},\\
U_{\rm AS} &= \frac{e^2}{4\pi \epsilon_r \epsilon_0}\int {\mathrm d}{\bf r}_1 \!\int{\mathrm d}{\bf r}_2 \,\frac{|\psi_{\rm AS}({\bf r}_1)|^2 |\psi_{\rm AS}({\bf r}_2)|^2}{|{\bf r}_1 - {\bf r}_2|},
\end{split}
\end{equation}
where the screening introduced by $\epsilon_r = 2$ is justified, even for an isolated molecule, by the presence of a polarizable core electrons.
We have performed the integrals in Eq.~\eqref{Eq:dir_Interaction_integrals} using the DFT wave functions plotted in Fig.~1 of the main text, with
the help of a Montecarlo method. The very similar numerical results ($U_{\rm S} = 1.37 e$V, $U_{\rm AS} = 1.43 e$V, and  $U_{\rm SAS}= 1.37 e$V), together with the partial convergence of the Montecarlo method when applied to the DFT molecular orbitals suggested us to simplify the model to a single parameter $U = 1.4e$V.
The robustness of the level ordening in the many-body spectrum with respect to variations of the interaction parameters within the estimated error
given by the Montecarlo method has been tested. The exchange parameter $J$ has been similarly calculated from the formula:
\begin{equation}
\label{Eq:exc_Interaction_integral}
J = \frac{e^2}{4\pi \epsilon_r \epsilon_0}\int {\mathrm d}{\bf r}_1 \!\int{\mathrm d}{\bf r}_2 \,
\frac{\psi_{\rm AS}^*({\bf r}_1)\psi_{\rm S}^*({\bf r}_2) \psi_{\rm AS}({\bf r}_2)\psi_{\rm S}({\bf r}_1)}{|{\bf r}_1 - {\bf r}_2|},
\end{equation}
which yields $J = 0.75 e$V. Since the molecular orbitals are real functions, Eq.~\eqref{Eq:exc_Interaction_integral} also gives the pair hopping integral.

\noindent The remaining parameters ($\epsilon_\mathrm{S}$ and $\epsilon_\mathrm{AS}$ and $\delta$) are
fitted to experimental data from the d$I$/d$V$ measurements.
The position of the resonant peaks in the differential conductance curves corresponding to the transition $|N,m\rangle \leftrightarrow |N+1,n\rangle$ is
given by the relation:
\begin{equation}
eV_{\rm b,res}^{N,m \leftrightarrow N+1,n} = \frac{1}{\alpha_T}(E^\mathrm{G}_{N+1,n}-E^\mathrm{G}_{N,m}),
\end{equation}
where $\alpha_\mathrm{T}$ ,the voltage drop at the tip, is defined by $\mu_T = \mu_0 + \alpha_\mathrm{T} eV_{\rm b}$, and for the
the grand canonical energies $E^\mathrm{G}_{N,m}$ it holds $E^\mathrm{G}_{N,m} = E_{N,m} - \delta N^2 + \phi_0N$,
where $E_{N,m}$ is the eigenvalue of the $m$-th excited $N$ particle state of $\HH{mol}$ (analytical expressions for $E_{N,m}$
are given in Table~1. In the experimental data (Fig.~3 in the main text)
we identify 5 resonance biases $V_{\rm b,res}$ from which, besides $\epsilon_\mathrm{S}$ and $\epsilon_\mathrm{AS}$ and $\delta$, also the parameter $\alpha_T$
can be extracted.

\noindent We assign the resonances seen in the  measurements on Cu(111), $\phi_{111}=4\,e$V from low to high bias
to specific transitions between 0 and 1 particle states: $V_{111}^{>0}=V_\mathrm{b,res}^{0\leftrightarrow10} = 0.7$V and $V_{111}^{>1}=V_\mathrm{b,res}^{0\leftrightarrow11} = 1.1$V, yielding
\begin{equation}
\begin{split}
	eV_{111}^{>0}	&=	\alpha_\mathrm{T}^{-1}\left( \epsilon_\mathrm{S} - \delta + \phi_{111} \right),\\
	eV_{111}^{>1}	&=	\alpha_\mathrm{T}^{-1}\left( \epsilon_\mathrm{S}+\Delta - \delta + \phi_{111} \right).
\end{split}
\end{equation}
where $\Delta = \epsilon_\mathrm{AS} - \epsilon_\mathrm{S}$. Analogously, we assign the resonances in the d$I$/d$V$
measurements on Cu(311), $\phi_{311}=3\,$eV. The only negative bias resonance is associated to a transition between 1 and 2 particle states: $V_{311}^{<0}=V_\mathrm{b,res}^{1\leftrightarrow20} = -0.7$V. The ones at positive bias involve instead 2 and 3 particle states: $V_{311}^{>0}=V_\mathrm{b,res}^{2\leftrightarrow30} = 0.9$V, and $V_{311}^{>1}=V_\mathrm{b,res}^{2\leftrightarrow31} =1.3$V, we get
\begin{equation}
\begin{split}
	eV_{311}^{<0}	&=	\alpha_\mathrm{T}^{-1}\left( \epsilon_\mathrm{S}+\Delta+U-\sqrt{\Delta^2+J^2}-3\delta+\phi_{311} \right),	\\
	eV_{311}^{>0}	&=	\alpha_\mathrm{T}^{-1}\left( \epsilon_\mathrm{S}+2U-J+\sqrt{\Delta^2+J^2}-5\delta+\phi_{311} \right),		\\
	eV_{311}^{>1}	&=	\alpha_\mathrm{T}^{-1}\left( \epsilon_\mathrm{S}+\Delta+2U-J+\sqrt{\Delta^2+J^2}-5\delta+\phi_{311} \right).
\end{split}
\end{equation}
where $e$ is the elementary charge taken with positive sign. It is now straightforward to determine the bias drop and the parameters of the Hamiltonian: $\alpha_\mathrm{T}	=	0.70$,	
$\delta = 0.43\,\mathrm{eV}$, $\epsilon_\mathrm{S}=-3.08\,\mathrm{eV}$, and $\epsilon_\mathrm{AS} =-2.8\,\mathrm{eV}$.

\begin{table}
     \begin{tabular}{llllll}
\multicolumn{6}{l}{Table 1. Many-body eigenenergies $E_{N,m}$ of $\HH{mol}$, omitting the spin degrees of freedom} \\

     \toprule
	\mc{$m\backslash N$}    &	\mc{0}	&	\mc{1}							&	\mc{2}												&	\mc{3}								&	\mc{4}	\\									 
    \midrule
	\mc{0}					&	0		&	$\epsilon_\mathrm{S}$			&	$2\epsilon_\mathrm{S}+\Delta+U-\sqrt{\Delta^2+J^2}$	&	$3\epsilon_\mathrm{S}+\Delta+3U-J$	&	 $4\epsilon_\mathrm{S}+2\Delta+6U-2J$	\\
	\mc{1}					&			&	$\epsilon_\mathrm{S}+\Delta$	&	$2\epsilon_\mathrm{S}+\Delta+U-J$					&	$3\epsilon_\mathrm{S}+2\Delta+3U-J$	&											 \\
	\mc{2}					&			&									&	$2\epsilon_\mathrm{S}+\Delta+U+J$					&										&											 \\
	\mc{3}					&			&									&	$2\epsilon_\mathrm{S}+\Delta+U+\sqrt{\Delta^2+J^2}$	&										&											 \\
    \bottomrule
   \end{tabular}
\end{table}

\section{Dynamics and transport}

\noindent The transport characteristics for the STM single molecule junction with thin insulating film, are obtained following
the approach already introduced by some of the authors~\cite{Sobczyk2012,Donarini2012,Siegert2013} in earlier works.
We summarize here only the main steps of the calculation. The junction is described by the Hamiltonian
$\HH{}=\HH{G} + \HH{S} + \HH{T} + \HH{tun}$, where, beside the grand canonical Hamiltonian $\HH{G}$ for the
molecule, $\HH{S}$ and $\HH{T}$ correspond to substrate (S) and tip (T), respectively and $\HH{tun}$
contains the tunnelling dynamics. The tip and the substrate are treated as noninteracting electronic leads:
\begin{align}
 \hat{H}_{\eta=\mathrm{S,T}}=\sum_{\kk\sigma}\eps{\eta\kk}\,\erz{c}{\eta\kk\sigma}\ver{c}{\eta\kk\sigma},
\end{align}
 where $\erz{c}{\eta\kk\sigma}$ creates an electron in lead $\eta$ with spin $\sigma$ and momentum $\kk$.
The tunneling Hamiltonian $\HH{tun}$ is given by
\begin{align}
 \HH{tun} = \sum_{\eta\kk i\sigma} t^\eta_{\kk i}\, \erz{c}{\eta\kk\sigma}\ver{d}{i\sigma} +\mathrm{h.c.},
\end{align}
and it contains the tunneling matrix elements $t^\eta_{\kk i}$, which are obtained by calculating
the overlap between the lead wavefunctions $\ket{\eta\kk}$ and the molecular orbitals $\ket{i}$~\cite{Sobczyk2012}.
The latter are the starting point for the calculation of the single particle tunnelling rate matrices $\Gamma^\eta_{ij}(E)=\frac{2\pi}{\hbar}\sum_\kk\, t^\eta_{i\kk}\left(t^\eta_{j\kk}\right)^*\delta\left(\eps{\eta\kk}-E\right)$ and, eventually, of the
many-body rates:

\begin{equation}
\label{Eq:ManyBody_Rates}
\begin{split}
R^{N,n\to N+1,m}_{\sigma\eta}  &= \sum_{ij}\Gamma^\eta_{ji}(E^G_{N+1,m}-E^G_{N,n})\\
&\langle N+1,m | \erz{d}{i\sigma}| N,n \rangle \langle N,n| \ver{d}{j\sigma} | N+1,m\rangle f^+(E^G_{N+1,m}-E^G_{N,n}-\alpha_\eta e V_\mathrm{bias},T)\\
R^{N,n\to N-1,m}_{\sigma\eta}  &= \sum_{ij}\Gamma^\eta_{ij}(E^G_{N-1,m}-E^G_{N,n})\\
&\langle N-1,m | \ver{d}{i\sigma}| N,n \rangle \langle N,n| \erz{d}{j\sigma} | N-1,m\rangle f^-(E^G_{N,n}-E^G_{N-1,m}-\alpha_\eta e V_\mathrm{bias},T),\\
\end{split}
\end{equation}
where $f^+(E) = (1+\exp(\beta E))^{-1}$ is the Fermi distribution with $\beta = (k_\mathrm{B}T)^{-1}$ and $f^-(E) =1 - f^+(E)$. Eq. \eqref{Eq:ManyBody_Rates} clearly show how each manybody rate is in general the superposition of several molecular orbitals, whose population is changed by the creation (annihilation) operator $\erz{d}{i\sigma}$ ($\ver{d}{i\sigma}$).

\noindent The system dynamics is calculated by means of the generalized master equation,
\begin{align}\label{GME}
 \dot{\rho}_\mathrm{red} = \mathcal{L}[\rho_\mathrm{red}],
\end{align}
for the reduced density operator~\cite{Darau2009,Sobczyk2012}
$\rho_\mathrm{red}=\operatorname{Tr}_{\mathrm{S,T}}\left(\rho\right)$.
The Liouvillian superoperator in Eq. \eqref{GME}
\begin{align}
 \mathcal{L} = \mathcal{L}_\mathrm{S} + \mathcal{L}_\mathrm{T} + \mathcal{L}_\mathrm{rel}
\end{align}
contains the terms $\mathcal{L}_\mathrm{S}$ and $\mathcal{L}_\mathrm{T}$ describing tunneling from and to the substrate and the tip, respectively.
These superoperators are combinations\cite{Sobczyk2012} of the many body rates in Eq.~(\ref{Eq:ManyBody_Rates}).
To account for relaxation processes independent from the electron tunnelling, similarly to Ref.~\cite{Koch2005},
we included the term $\mathcal{L}_{\mathrm{rel}}$:
\begin{align}\label{L_rel}
  \mathcal{L}_\mathrm{rel}\left[\rho\right] = -\frac{1}{\tau}\left( \rho - \sum_N\rho^{\mathrm{th},N} \sum_l\, \rho^N_{ll} \right).
\end{align}
$\mathcal{L}_{\mathrm{rel}}$ combines different relaxation processes associated e.g. with  the phonon emission or with particle-hole excitation
in the substrate within the relaxation time approximation. This term induces the relaxation of each N-particle subblock of $\rho$ towards its 
(canonical) thermal distribution:
\begin{align}
 \rho^\mathrm{th,N}=\sum_{k}\, \frac{ \ee{-\beta E_{Nk} } }{ \sum_l \ee{-\beta E_{Nl} } } \ket{Nk}\bra{Nk},
\end{align}
with $\beta=\left(k_\mathrm{B}T\right)^{-1}$. The speed of the process is set by the relaxation time $\tau$. 
Since $\mathcal{L}_\mathrm{rel}$ acts separately on each $N$-particle subblock, it conserves the particle number on the molecule and thus does not contribute directly to the transport.
For the calculation of the long time dynamics, we are interested in the stationary solution $\rho_\mathrm{red}^\infty$ for which $\dot\rho_\mathrm{red}^\infty=\mathcal{L}[\rho_\mathrm{red}^\infty]=0$.
Eventually, the stationary current through the system is evaluated as
\begin{align}\label{Eq_current}
\braket{\hat I_\mathrm{\eta}}=\operatorname{Tr}_\mathrm{mol}\left(\hat N\mathcal{L_\eta}[\rho_\mathrm{red}^\infty]\right),
\end{align}
being $\hat I_\eta=\hat N\mathcal{L_\eta}$ the current operator for the lead $\eta$.

\section{Level alignment}

\noindent In previous studies of molecules on insulating films it was observed that, due to the electronic decoupling by the film, the molecular levels are roughly aligned with the vacuum level. From an electrochemical characterization the electron affinity of DCV5T in solution was determined to be at $-3.73$\,eV relative to the vacuum level~\cite{Fitzner2011}. The polarizability of the solution lowers the electron affinity level, such that here the LUMO transport level can be expected at some tenths of an eV higher in energy. 
Considering the work function of NaCl/Cu(111) of about  $4\,$eV~\cite{Repp2005,Bennewitz1999}, this expectation in good agreement with the experimentally observed position of the S state for this system.

\section{Kelvin probe force spectroscopy measurements}

\begin{figure*}
  \centering
  \includegraphics[width=17 cm]{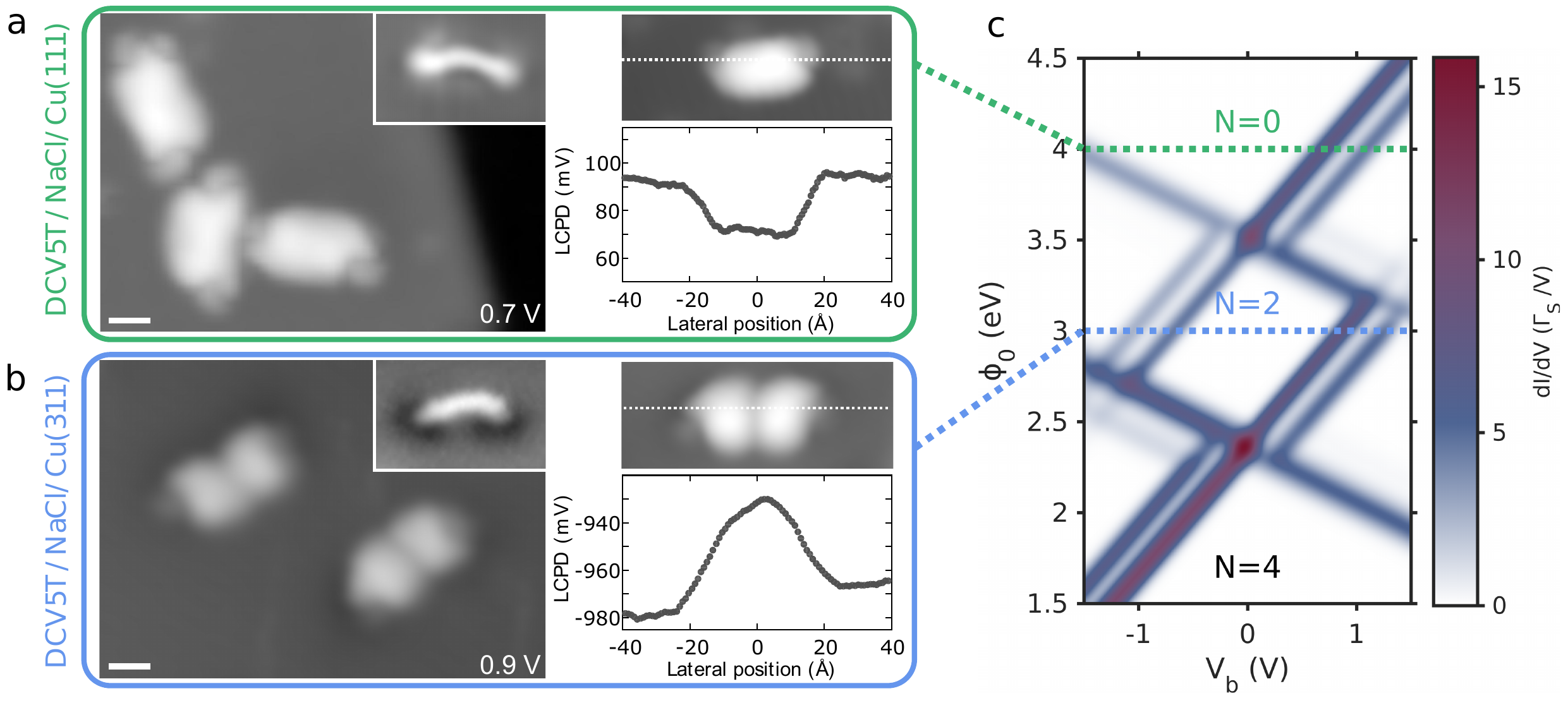}
  \caption{(a and b, left) STM images of the first DCV5T electronic resonance for NaCl/Cu(111) (a) and NaCl/Cu(311) (b) as substrates 
 (Insets show corresponding STM images at voltages below the first molecular resonance). Scale bar 1 nm.
 (a and b, right) Local contact potential difference across the molecule 
 for NaCl/Cu(111) (a) and NaCl/Cu(311) (b) as substrates.
(c) Stability diagram for DCV5T as a function of substrate work function and bias across the junction. The equilibrium particle numbers are indicated within the low conductance diamonds. The upper and lower dashed lines correspond to the NaCl/Cu(111) and NaCl/Cu(311) substrates, respectively. }
   \label{SI1}
\end{figure*}

\noindent We performed Kelvin probe force spectroscopy (KPFS) measurements along the molecules for both substrates, as is shown in Fig.~\ref{SI1}. From a fit to the parabolic shape of the frequency shift $\Delta f(V)$ as a function of sample voltage $V$, the local contact potential difference (LCPD) between tip and sample\cite{Nonnenmacher1991, Sadewasser2009, Mohn2012, Steurer2015} is extracted. Next to the molecules, on the clean NaCl films, the LCPD differs by slightly more than $1$\,eV for the two systems providing a rough estimate of the work function difference for the two systems \footnote{For the two KPFS measurements of the two different systems the tip apex condition was not necessarily identical.} in accordance with literature values~\cite{Gartland1972,Repp2005,Olsson2005}.
Since local surface charges and dipoles affect the LCPD above adsorbates, the latter should qualitatively reflect the charge state~\cite{Gross2009}, the electron affinity~\cite{Schuler2014}, and the charge distribution~\cite{Neff2015,Mohn2012,Albrecht2015}. The decrease of about 20\,meV in LCPD over the molecule in the case of DCV5T/NaCl/Cu(111) we assume to be due to the large electron affinity of DCV5T. On the NaCl/Cu(311) substrate the observed increase of LCPD is consistent with an anionic state of DCV5T/NaCl/Cu(311)~\cite{Ikeda2008}.


\end{document}